\documentclass[%
 reprint,
 amsmath,amssymb,
 aps,
 floatfix,
 longbibliography,
]{revtex4-2}

\usepackage{graphicx}
\usepackage{bm}
\usepackage[utf8]{inputenc}
\usepackage[T1]{fontenc}
\usepackage[%
  colorlinks=true,
  urlcolor=blue,
  linkcolor=blue,
  citecolor=blue]{hyperref}
\usepackage[all]{hypcap}
\usepackage{sidecap}
\sidecaptionvpos{figure}{t}
\usepackage[capitalise]{cleveref}
\graphicspath{{pictures/}}

\begin{document}

\title{Spin-polarized two-dimensional electron/hole gas at the interface of non-magnetic semiconducting 
half-Heusler compounds: Modified Slater-Pauling rule for half-metallicity at the interface}

\author{Emel G\"{u}rb\"{u}z$^{1}$}        
\author{Sukanya Ghosh$^{1}$} 
\author{Ersoy \c{S}a\c{s}{\i}o\u{g}lu$^{2}$}\email{ersoy.sasioglu@physik.uni-halle.de}
\author{Iosif Galanakis$^{3}$}\email{galanakis@upatras.gr}
\author{Ingrid Mertig$^{2}$}
\author{Biplab Sanyal$^{1}$}\email{biplab.sanyal@physics.uu.se}

\affiliation{$^{1}$Department of Physics and  Astronomy,  Uppsala  University,  75120  Uppsala,  Sweden \\
$^{2}$Institute of Physics, Martin Luther University Halle-Wittenberg, 06120 Halle (Saale), Germany \\
$^{3}$Department of Materials Science, School of Natural Sciences, University of Patras, GR-26504 Patra, Greece}

\date{\today}

\begin{abstract}

Half-Heusler compounds with 18 valence electrons per unit cell are well-known 
non-magnetic semiconductors. Employing first-principles electronic band structure 
calculations, we study the interface properties of the half-Heusler heterojunctions
based on FeVSb, CoTiSb, CoVSn, and NiTiSn compounds, which belong to this category 
of materials. Our results show that several of these heterojunction interfaces become 
not only metallic but also magnetic. The emergence of spin-polarization is accompanied 
by the formation of two-dimensional electron gas (2DEG) or hole gas (2DHG) at the 
interface. We qualitatively discuss the origin of the spin polarization at the interfaces 
on the basis of the Stoner model. For the cases of magnetic interfaces where 
half-metallicity is also present, we propose a modified Slater-Pauling rule similar 
to the one for bulk half-metallic half-Heusler compounds. Additionally, we calculate 
exchange parameters, Curie temperatures and magnetic anisotropy energies for magnetic 
interfaces. Our study, combined with the recent experimental evidence for the presence 
of 2DEG at CoTiSb/NiTiSn heterojunctions might motivate future efforts and studies 
toward the experimental realization of devices using the proposed heterojunctions.

\end{abstract}


\maketitle

\section{Introduction}\label{sec1}

Heusler compounds, named after Fritz Heusler \cite{Heusler1903,Heusler1912}, 
are ternary and quaternary intermetallic compounds that crystallize in 
close-packed lattice structures \cite{Graf2011,Tavares2023,Chatterjee2022}.
Especially, the discovery of half-metallicity (the electronic 
band structure is metallic for one spin channel and semiconducting for the other 
\cite{Katsnelson2008}) in Heusler compounds, which was followed by other exotic behaviors like 
spin-gapless semiconductors and spin-filter materials \cite{Galanakis2016},
led to the proposal of novel devices \cite{Sasioglu2019,Sasioglu2020,Kawasaki2019,Aull2019,Aull2021}.
An important role in the rapid growth of this research field was played by the 
first-principles electronic band structure calculations. On one hand, they successfully explained the 
origin of half-metallicity and linked it to the magnetic properties through 
the so-called Slater-Pauling rules \cite{Galanakis2002a,Galanakis2002b,Galanakis2004,Skaftouros2013,Ozdogan2013,Nepal2022}, 
and on the other hand, extended databases built using such calculations resulted in 
the prediction of hundreds of new Heusler compounds which were later grown experimentally \cite{Ma2017,Gillessen2009,Gillessen2010,Faleev2017a,Faleev2017b,Faleev2017c,Gao2019,Han2019}.

Heusler compounds are categorized into various families depending on the number of 
atoms in the unit cell and their ordering \cite{Graf2011,Tavares2023}. The ones 
having the chemical formula XYZ, where X and Y are transition-metal atoms and Z is 
a metalloid is named half-Heusler (or semi-Heusler compounds). When the number 
of valence electrons in the unit cell exceeds 19 and goes up to 22, most of them 
are half-metals \cite{Galanakis2002a}. As shown by Galanakis \textit{et al.} the 
half-metallicity is directly connected to the total spin magnetic moment through 
the Slater Pauling rule $M_\mathrm{t}=Z_\mathrm{t}-18$, where $M_\mathrm{t}$ is 
the total spin magnetic moment in the unit cell expressed in $\mu_\mathrm{B}$ and 
$Z_\mathrm{t}$ is the total number of valence electrons in the unit cell. 
The number 18 expresses the fact that there are exactly nine occupied states in 
the minority-spin electronic band structure, which exhibits semiconducting 
behavior. There is a single $s$ and a triple $p$ band low in energy stemming from 
the Z atom. The $d$ valence orbitals of the X and Y atoms hybridize creating five 
occupied bonding orbitals, which are separated by an energy gap from the five 
unoccupied antibonding orbitals.

The Slater-Pauling rule correctly predicts that half-Heusler compounds with 
exactly 18 valence electrons should be non-magnetic semiconductors with a gap 
in both spin-channels \cite{Galanakis2002a}. This ‘‘18-electron rule’’ for 
semiconducting half-Heusler compounds was also derived by Jung \textit{et al.} 
based on ionic arguments \cite{Jung2000}. Among the 18-valence electron 
half-Heusler compounds, CoTiSb, NiTiSn, FeVSb, and CoVSn have attracted most 
of the attention. Pierre and collaborators in 1994 have confirmed experimentally 
the non-magnetic semiconducting character of NiTiSn \cite{PIERRE1994}. Tobola \textit{et al.} 
have shown experimentally that CoTiSb is also a non-magnetic semiconductor \cite{Tobola1998}. 
The experimental findings for both NiTiSn and CoTiSb have been also confirmed 
by \text{ab-initio} calculations in Ref. \cite{Tobola1998}. Recently, Ouardi 
\textit{et al.} have synthesized CoTiSb and investigated it both theoretically and 
experimentally \cite{Ouardi2012}. Lue and collaborators grew samples of CoVSn and 
their findings were consistent with a non-magnetic semiconducting behavior \cite{Lue2001}. 
Finally, Mokhtari and collaborators have  shown theoretically that FeVSb is also a 
non-magnetic semiconductor \cite{Mokhtari2018} followed by the experimental observation 
in 2020 by Shourov \textit{et al.} \cite{Shourov2020}. Ma \textit{et al.} in 2017 
studied using first-principles calculations a total of 378 half-Heusler compounds \cite{Ma2017}.
Among them, there were 27 compounds with 18 valence electrons, including the aforementioned 
ones, which were all found to be non-magnetic semiconductors \cite{Ma2017}. Doping 
these compounds with transition-metal atoms 
\cite{Kroth2006,Naghibolashrafi2020,Nanda2005,Sanyal2006,Sun2015,Tareuchi2005,Tobola2000,Tobola2003,Balke2008} 
or vacancies \cite{Zhu2011} leads to a half-metallic behavior.

Recently, Sharan \textit{et al.} have studied employing \textit{ab-initio} calculations 
the formation of a two-dimensional electron gas (2DEG) or hole gas (2DHG) at the interface 
between CoTiSb and NiTiSn compounds \cite{Sharan2019}. To model the heterojunction, they 
assumed a superlattice along the [001] direction. Along the [001] direction,  CoTiSb is non-polar 
and NiTiSn is polar \cite{Sharan2019}. Similarly to the complex oxides' polar/non-polar 
interfaces \cite{Betancourt2017,Ohtomo2002,Ohtomo2004}, a 2DEG is formed at the 
(TiSb)-Ni interface while a 2DHG is formed at the Co-(TiSn) interface \cite{Sharan2019}. 
A 2DEG(2DHG) is a type of electronic system in which a large number of electrons (holes) 
are confined to a very thin, two-dimensional layer like the one occurring at an interface 
of a heterojunction. Electrons (holes) are free to move in the two dimensions, 
but are strongly confined in the third dimension leading to many potential applications, 
including high-speed electronic devices and quantum computers \cite{Linh1983}. The 2DEG 
should not be confused with the 2D electron liquid observed at surfaces of bulk 
semiconducting Heusler compounds \cite{Keshavarz2020}. Experimentally, Harrington has 
grown heterostructures made up of alternating 25nm thick CoTiSb and NiTiSn \cite{Harrington2018}.
Although the structure of the interface has not been studied, interface transport 
measurements suggest that 2DEG is present at the interface giving indirect evidence 
for its formation \cite{Harrington2018}.

The goal of this study is to provide a comprehensive understanding of the  electronic
and magnetic properties of interfaces formed by various combinations of CoTiSb, NiTiSn, 
FeVSn, and CoVSn non-magnetic semiconducting Heusler compounds along the [001] growth 
direction by employing state-of-the-art first-principles electronic band structure calculations. 
We find that for all heterojunctions except for CoTiSb/CoVSn, the emergence of 2DEG or 2DHG  
at the interfaces is accompanied by the occurrence of magnetism. We qualitatively discuss 
the origin of interface magnetism on the basis of the Stoner model. In some cases, our 
calculations suggest that also half-metallicity is present and we formulate a modified 
version of the Slater-Pauling rule to connect the magnetic properties to half-metallicity 
and the total number of valence electrons at the specific interface. Finally, we present 
the exchange constants for the magnetic interfaces and use them to predict the Curie 
temperature, which is important for applications. We should mention at this point that 
according to the Mermin-Wagner theorem,  the long-range magnetic order does not exist in 
one- (1D) or two-dimensional (2D) isotropic magnets. But as shown recently in Ref.\,\cite{Jenkins2020} 
short-range exchange interactions even in the absence of magnetic  anisotropy
can induce magnetic order in finite-size 2D magnets even for samples of millimeters size.  
Thus, the magnetic interfaces discussed in the present study are feasible and can
be realized in realistic spintronic devices. The rest of the manuscript is organized 
as follows: In Section\,\ref{sec2}  we present details of our calculations, in Section\,\ref{sec3} 
we present our results, and  finally, in Section\,\ref{sec4} we summarize our results and 
present the conclusions of our study.

\begin{figure*}[!t]
\centering
\includegraphics[width=0.99\textwidth]{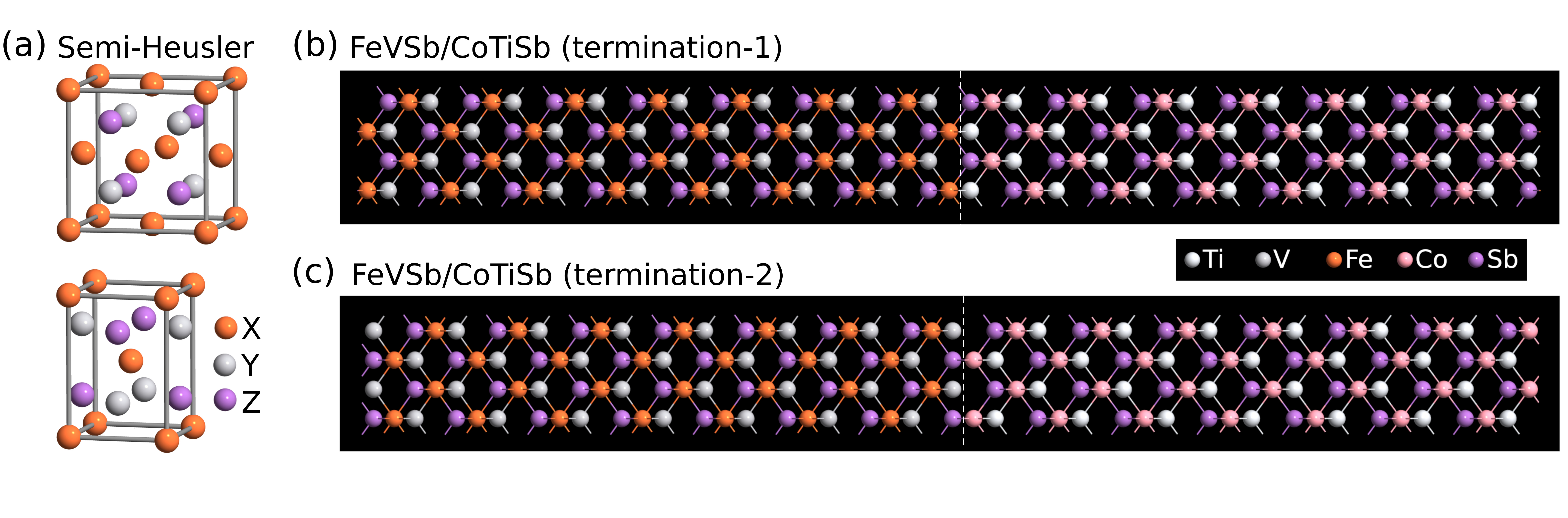}
\vspace{-0.8 cm}
\caption{(a) (Color online) Upper: Schematic representation of the conventional unit 
cell of XYZ half-Heusler compounds. Lower: Minimal tetragonal unit cell for the superlattice 
calculations (see text for details). (b) and (c) the structure of the superlattice for 
the two different possible interface terminations in the case of FeVSb/CoTiSb heterojunction.}
\label{fig1}
\end{figure*}

\section{Computational Method}\label{sec2}

The bulk half-Heusler compounds XYZ crystallize in the cubic $C1_\mathrm{b}$ lattice 
structures shown in the left panel of Fig.\,\ref{fig1}(a). The space group is the 
$F\overline{4}3m$ and actually consists of four interpenetrating f.c.c. sublattices; 
one is empty and the other three are occupied by the X, Y, and Z atoms. 
The unit cell is an f.c.c. one with three atoms as a basis along the long diagonal of 
the cube: X at $(0\:0\:0)$, Y at $(\frac{1}{4}\:\frac{1}{4}\:\frac{1}{4})$ and Z at 
$(\frac{3}{4}\:\frac{3}{4}\:\frac{3}{4})$ in Wyckoff coordinates. 
When we consider a superlattice along the [001] direction then the consecutive layers 
are made up of pure X and mixed YZ layers and the in-plane unit cell is a square with 
lattice parameter the $\frac{1}{\sqrt{2}}$ of the cube's lattice parameter as shown 
in the right panel of Fig.\,\ref{fig1}(a). For all four compounds FeVSb, CoTiSb, CoVSn, 
and NiTiSn we adopted the equilibrium lattice parameters calculated in Ref.\,\cite{Ma2017} 
using the Vienna Ab-initio Simulation Package (VASP) \cite{Kresse1994,Kresse1999} in 
conjunction with the generalized gradient approximation (GGA) to the exchange-correlation 
potential \cite{Perdew1996}. We present the lattice parameters of the bulk compounds 
adopted in our study in Table\,\ref{table1}. 

To carry out the spin-polarized density functional theory (DFT) calculations we employ the 
\textsc{QuantumATK} software package ~\cite{QuantumATK,QuantumATKb}. We use linear 
combinations of atomic orbitals (LCAO) as a basis set together with norm-conserving 
PseudoDojo pseudopotentials~\cite{VanSetten2018} with the  Perdew-Burke-Ernzerhof 
(PBE) parametrization of the GGA functional \cite{Perdew1996}. For the determination
of the ground-state properties of the bulk compounds, we use a $15 \times 15 \times 
15$ Monkhorst-Pack $\mathbf{k}$-point grid, while for periodic supercell calculations 
a $20 \times 20 \times 2$  Monkhorst-Pack $\mathbf{k}$-point grid is adopted \cite{Monkhorst1976}.

When we form heterojunctions using FeVSb, CoTiSb, CoVSn, and NiTiSn compounds, we get 
six possible combinations and for each combination, there are two possible interface
terminations denoted as Termination-1 and Termination-2 in Table\,\ref{table3} and 
shown in Fig.\,\ref{fig1} using the FeVSb/CoTiSb as an example (the two possible 
interfaces are made up of Fe-TiSb and VSb-Co layers respectively). If one of the two 
interface terminations generates a 2DEG the other will be 2DHG as we will discuss when we present 
our results. To simulate the heterojunction we assume a supercell consisting of 29 
(27) layers of FeVSb and 27 (29) layers of CoTiSb for the first (second) termination. 
To construct the supercell, we fix for the first material, for instance in FeVSb/CoTiSb  
supercell, the lattice parameter of FeVSb to be the cubic one shown in Table\,\ref{table1}. 
For the second compound, CoTiSb in our example, we consider an in-plane lattice parameter 
the one of FeVSb and we relax the out-of-plane lattice parameter along the [001] direction; 
in Table \ref{table2} we present the $c/a$ ratio for the second compound of the heterostructure 
in the supercell.   

\begin{figure*}[t]
\centering
\includegraphics[width=0.9999\textwidth]{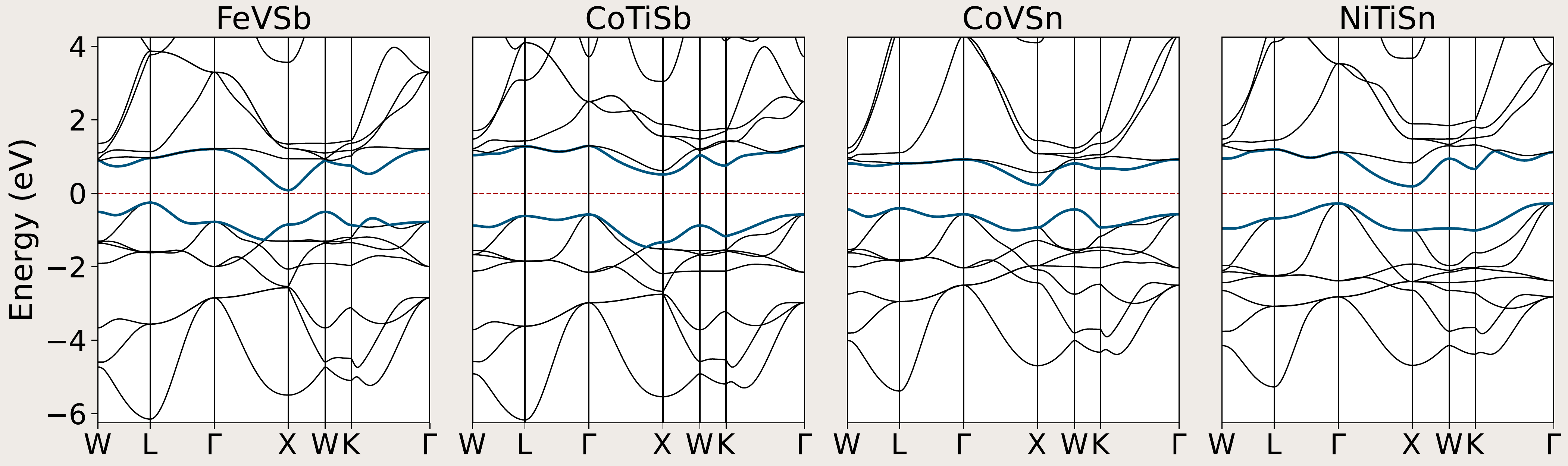}
\vspace{-0.4 cm}
\caption{(Color online) Band structure of the bulk FeVSb, CoTiSb, CoVSn, and NiTiSn 
half-Heusler compounds along the high-symmetry lines in the Brillouin zone. With blue we denote 
the topmost valence and lowest conduction bands. The Fermi level is set to zero energy.}
\label{fig2}
\end{figure*}

In the present study, we adopted the supercell approach instead of a two-terminal device 
model  implemented in QuantumATK and adopted in previous studies \cite{Sasioglu2019,Sasioglu2020,Aull2019,Aull2021} 
since the band structure calculations using the device model are computationally 
extremely demanding. For FeVSb/NiTiSn heterojunction, we performed also calculations 
using the device model with two semi-infinite leads by employing the nonequilibrium 
Green’s function (NEFG) approach combined with DFT. For these DFT+NEGF calculations, 
we used a  $20 \times 20 \times 115$ $\mathbf{k}$-point mesh. The results were identical 
to the ones obtained by the supercell approach and thus the use of the latter is completely 
justified.

To study finite-temperature properties of the interfaces we map the complex 
multi-sublattice itinerant electron problem onto a classical effective Heisenberg 
Hamiltonian
\begin{equation}
\label{eqn:Heisenberg}
    H_{\mathrm{eff}} = -\sum_{\mathrm{i,j}}\sum_{\mathrm{\mu,\nu}} J^{\mu\nu}_{\mathrm{ij}} 
    \bm S^{\mu}_i \cdot \bm S^{\nu}_j, 
\end{equation}
where $\mu$ and $\nu$ denote different sublattices at the interface, $i$ and $j$ indicate atomic 
positions, and $\bm S^{\mu}_i$ is the unit vector of the $i$ site in the $\mu$ sublattice. The 
Heisenberg exchange constants $J^{\mu\nu}_{\mathrm{ij}}$ are calculated by employing the Liechtenstein
formalism~\cite{liechtenstein1987local} within the full-potential linear muffin-tin orbital (FP-LMTO) 
code RSPt \cite{Wills201}. The crystalline structure information for the studied interfaces obtained 
with the LCAO is used as input for the  electronic structure calculations by the FP-LMTO  approach. 
According to our tests, both QuantumATK and FP-LMTO methods provide a very similar electronic 
structure for the systems under study. Note that in the calculation of exchange parameters, we 
take into account atoms at the two interface layers and the two sub-interface layers only.

To estimate the Curie temperature $T_C$ of the magnetic interfaces, we use the mean-field approximation 
for a multi-sublattice system~\cite{anderson1963theory,Ersoy_2004,Yamada2018}, which is given by
\begin{equation}
\label{eqn:TcRPA}
T_C=\frac{2}{3k_B}J_{\mathrm{L}}^{\mu\nu},
\end{equation}
where $J_{\mathrm{L}}^{\mu\nu}$ is the largest eigenvalue of $J_{0}^{\mu\nu}=\sum_{j}J^{\mu\nu}_{0\mathrm{j}}$.

\section{Results and Discussion}\label{sec3}

Prior to presenting our results, we should comment on our choice to use PBE which is a GGA
functional in our study. GGA functionals are well-known to underestimate the band gap of 
semiconductors and to this respect, more elaborated functionals like the hybrid ones, which 
are much more demanding in computer sources, have been developed. The latter are 
semi-empirical combining the exact Hartree-Fock exchange with GGA and accurately reproducing 
the energy band gaps in usual semiconductors \cite{Garza2016}. In Ref. \cite{Sharan2019}, 
authors employed such a hybrid functional; the so-called Heyd, Scuseria, and Ernzerhof 
hybrid functional (HSE06) \cite{Heyd03,Heyd06} as implemented in the VASP code \cite{Kresse1994,Kresse1999}. 
They calculated for NiTiSn an energy band gap of 0.65 eV and for CoTiSb a value of 1.45 eV. 
These values are considerably larger than the 0.44 eV and 1.06 eV, respectively, calculated 
using PBE in Ref.\,\cite{Ma2017} (as we will later discuss our PBE values are in excellent 
agreement with these values although we used a different electronic structure code). But, 
Sharon and collaborators have ignored that Heusler compounds like CoTiSb and NiTiSn are not 
usual semiconductors since they contain transition metal atoms and thus the accuracy of the 
semi-empirical hybrid functionals is not granted. 

To clarify the above point we have to compare the above \textit{ab-initio} calculations with 
experimental results. In Ref.\,\cite{Ouardi2012} Ouardi and collaborators have determined the 
energy band gap of CoTiSb both experimentally as well as using ab-initio calculations employing 
the PBE functional. Their calculations have shown that CoTiSb exhibits an indirect gap of 
1.06 eV, identical to the PBE-derived value in Ref.\,\cite{Ma2017},  in excellent
agreement with their experimental value of about 1.0 eV. Moreover, calculations produced 
an optical gap (direct gap at the $\Gamma$ point) of 1.83 eV also in perfect agreement with
 the experimental value of 1.8 eV. Finally, the experimentally determined lattice constant
 was 5.884 \AA\ in perfect agreement with their own PBE calculations as well as the PBE 
 calculations in Ref.\,\cite{Ma2017}.  Thus, one can safely conclude that the PBE functional 
 is the appropriate one to study semiconductors like the half-Heusler compounds under study  
 and the  semi-empirical hybrid functionals overestimate the band gap in these materials. 
 Moreover,  as was shown in the case of  transition-metal-based  full-Heusler semiconductors, 
 many-body correlations calculated using the \textit{GW} approximation has minimal effect on 
 the PBE calculated electronic band structure and the energy gaps increase by less than 
 0.2 eV in all cases \cite{Tas2016}. We should finally note that, as shown in Ref.\,\cite{Haas2009}, 
 the use of the more elaborated meta-GGA functionals for semiconductors lead to energy gaps 
 that are less accurate with respect to the PBE calculated  ones when compared to experimental data. 
.

\subsection{Electronic structure  of bulk semiconducting half-Heusler compounds}

The starting point of our study is the calculation of the electronic properties of bulk 
materials. As mentioned above for all four compounds FeVSb, CoTiSb,  CoVSn, and NiTiSn, 
we have adopted the equilibrium lattice constants calculated in Ref.\,\cite{Ma2017} and 
we present them in Table\,\ref{table1} together with the band gap values. All four 
compounds were found to be non-magnetic semiconductors in agreement with previous 
first-principles calculations \cite{PIERRE1994,Tobola1998,Lue2001,Mokhtari2018,Ma2017} 
and in Fig.\,\ref{fig2} we present the calculated electronic band structure along the
high-symmetry directions in the Brillouin zone. As the band structure plots reveal, 
they are indirect  band-gap semiconductors but the valence band maximum (VBM) and the 
conduction band  minimum (CBM) do not occur at the same  high-symmetry points for all 
compounds. The character of the bands follows the discussion on the minority-spin band 
structure in half-Heusler compounds \cite{Galanakis2002a}. There are exactly nine occupied 
bands below the Fermi level and each band accommodates two electrons due to the spin-degeneracy. 
The lowest band, not-shown in Fig.\,\ref{fig2} stems from the $s$ states of the Sn(Sb) 
atoms. The lowest shown bands, which are triply degenerate at the $\Gamma$ point come 
from the valence $p$ bands of the Sn(Sb) atoms. Afterward, there are two almost flat bands, 
which are degenerate at the $\Gamma$ point and stem from the bonding $d$ states between 
the transition metal atoms. These bonding orbitals are of $e_\mathrm{g}$ character. 
Just below the Fermi level, the bands are triply degenerate at the $\Gamma$ point and 
stem from the bonding $t_\mathrm{2g}$ orbitals between the neighboring transition metal 
atoms. These last are separated with a gap from the antibonding  $d$ states stemming 
from the hybridization between the $d$ states of the transition-metal atoms.

\begin{table}[b]
\caption{Equilibrium lattice parameters (a) taken from Ref.\cite{Ma2017}, number of valence 
electrons per unit cell ($Z_\mathrm{t}$), band gap $E_\mathrm{g}$, position of the  
valence band maximum ($E_{VB}$) and of the conduction band minimum ($E_{CB}$) with 
respect to the Fermi level for FeVSb, CoTiSb, CoVSn, and NiTiSn  half-Heusler compounds.}
\begin{ruledtabular}
\begin{tabular}{lccccc}
Compound &  a(\AA)  & $Z_t$  &  $E_{\mathrm{g}}$(eV)  &  $E_{\mathrm{VB}}$(eV)& $E_{\mathrm{CB}}$(eV) \\
\hline
FeVSb   & 5.78 & 18 & 0.34 & -0.26 & 0.08  \\
CoTiSb  & 5.88 & 18 & 1.09 & -0.58 & 0.51 \\
CoVSn   & 5.79 & 18 & 0.63 & -0.41 & 0.22 \\
NiTiSn  & 5.93 & 18 & 0.47 & -0.28 & 0.19 \\
\end{tabular}
\label{table1}
\end{ruledtabular}
\end{table}

\begin{table*}[t]
\caption{Six possible half-Heusler heterojunctions made up of the four considered 
compounds. The $c/a$ ratios  are provided for the second half-Heusler compound in 
the heterojunction. For the two layers at each interface, we provide the calculated 
atomic and total spin magnetic moments as well as the ideal value for half-metallicity 
predicted by the $M_{\mathrm{t}}^{\mathrm{SP}} = \frac{Z_{\mathrm{t}}}{2}-9$ Slater-Pauling 
rule. The last column describes the character of each interface termination with respect 
to the 2-dimensional electron gas (2DEG). SPHG stands for spin-polarized hole gas, 
HG for hole gas, SPEG for spin-polarized electron gas, and EG for electron gas.}
\begin{ruledtabular}
\begin{tabular}{@{}*{1}l*{15}{l}@{}}
 Half-Heusler      &  $c/a$     &  Termination-1 & &   &    &       &  &     &  Termination-2 &    & & &     &       &       \\
 Heterojunction         &          &  Composition   & $Z_{\mathrm{t}}$ & $M_{\mathrm{X}}$  & $M_{\mathrm{Y}}$  & $M_{\mathrm{t}}$ & $M_{\mathrm{t}}^{\mathrm{SP}}$ & 2DEG &   Composition  & $Z_{\mathrm{t}}$ &   $M_{\mathrm{X}}$  & $M_{\mathrm{Y}}$  & $M_{\mathrm{t}}$ & $M_{\mathrm{t}}^{\mathrm{SP}}$ & 2DEG  \\  
\hline
FeVSb/CoTiSb   & 1.03   &  Fe|TiSb & 17 & -0.49 & 0.08 & -0.42 & -0.5 & SPHG  & VSb|Co  & 19 & -0.10 & 0.76 & 0.64 & 0.5 & SPEG  \\
FeVSb/CoVSn    & 1.01   &  Fe|VSn  & 17 & -0.81 & 0.29 & -0.54 & -0.5 & SPHG  & VSb|Co  & 19 & -0.03 & 0.18 & 0.15 & 0.5 & SPEG  \\
FeVSb/NiTiSn   & 1.05   &  Fe|TiSn & 16 & -1.12 & 0.17 & -0.99 & -1.0 & SPHG  & VSb|Ni  & 20 &  0.01 & 1.26 & 1.23 &1.0 & SPEG  \\
CoTiSb/CoVSn   & 0.98   &  Co|VSn  & 18 &  0.00 & 0.00 & 0.00  & 0.0 &   & TiSb|Co & 18 &  0.00 & 0.00 & 0.00 & 0.0 &     \\
CoTiSb/NiTiSn  & 1.02   &  Co|TiSn & 17 &  0.00 & 0.00 & 0.00 & -0.5 & HG    & TiSb|Ni & 19 &  0.00 & 0.00 & 0.00 & 0.5 & EG  \\
CoVSn/NiTiSn   & 1.05   &  Co|TiSn & 17 &  0.00 & 0.00 & 0.00 & -0.5 & HG    & VSn|Ni  & 19 & -0.01 & 0.58 & 0.53 & 0.5 & SPEG  \\
\end{tabular}
\label{table2}
\end{ruledtabular}
\end{table*}

In Table \ref{table1}, we also present the calculated band gap $E_\mathrm{gap}$ values, in eV 
units. The calculated values are in ascending order 0.34, 0.47, 0.63, and 1.09 eV for FeVSb, 
NiTiSb, CoVSn, and CoTiSb respectively. These values are very close to the ones calculated for 
the same lattice constants by Ma \textit{et al.} in Ref.\,\cite{Ma2017} (their values were 0.38, 
0.44, 0.65, and 1.06 respectively). This shows that the adopted electronic band structure method 
for the calculations is not crucial to calculate the properties and results depend strongly on 
the choice of the exchange-correlation functional; in both studies (ours and the study of Ma 
\textit{et al.}) the PBE parametrization of the GGA functional has been used). Moreover, our result 
for CoTiSb agrees well with the  experimental value of about 1 eV \cite{Ouardi2012}. Finally, 
in Table \ref{table1}, we have also included the relative position of the VBM and CBM with respect 
to the Fermi level calculated as the Fermi level minus the corresponding energy VBM position and 
this explains why the $E_\mathrm{VB}$ has a negative sign while $E_\mathrm{CB}$ has a positive sign.

\begin{figure}[!b]
\centering
\includegraphics[width=\columnwidth]{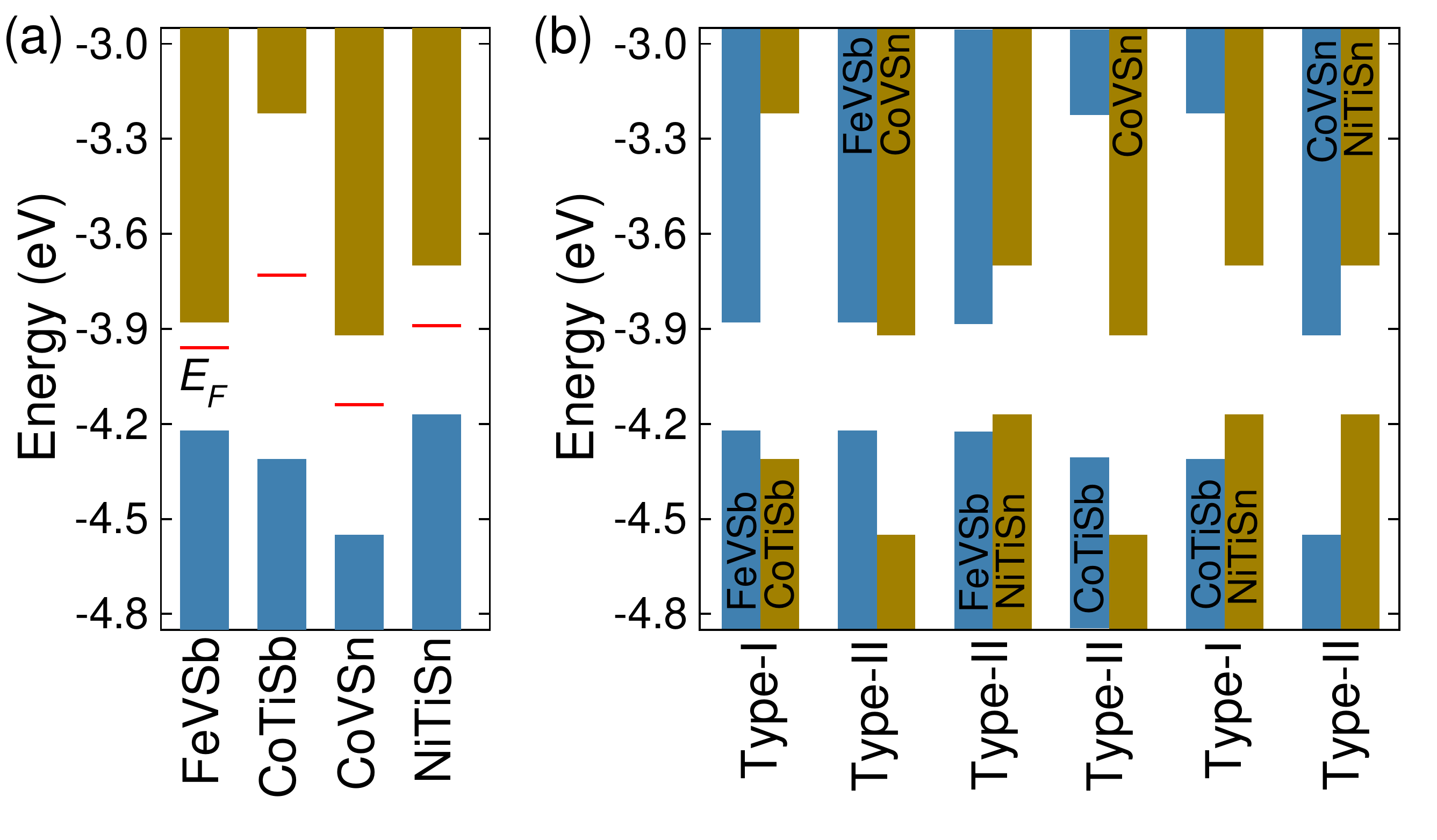}
\vspace{-0.3 cm}
\caption{(a) Position of the valence and conduction bands with respect to the Fermi level in 
bulk Heusler compounds. (b) Band alignment for the various interfaces under study. 
Type-I and type-II denote the two possible types of band alignment (see text for details).}
\label{fig3}
\end{figure}

\subsection{Spin-polarized 2DEG and 2DHG at the Interfaces}

\begin{figure*}[!ht]
\centering
\includegraphics[width=\textwidth]{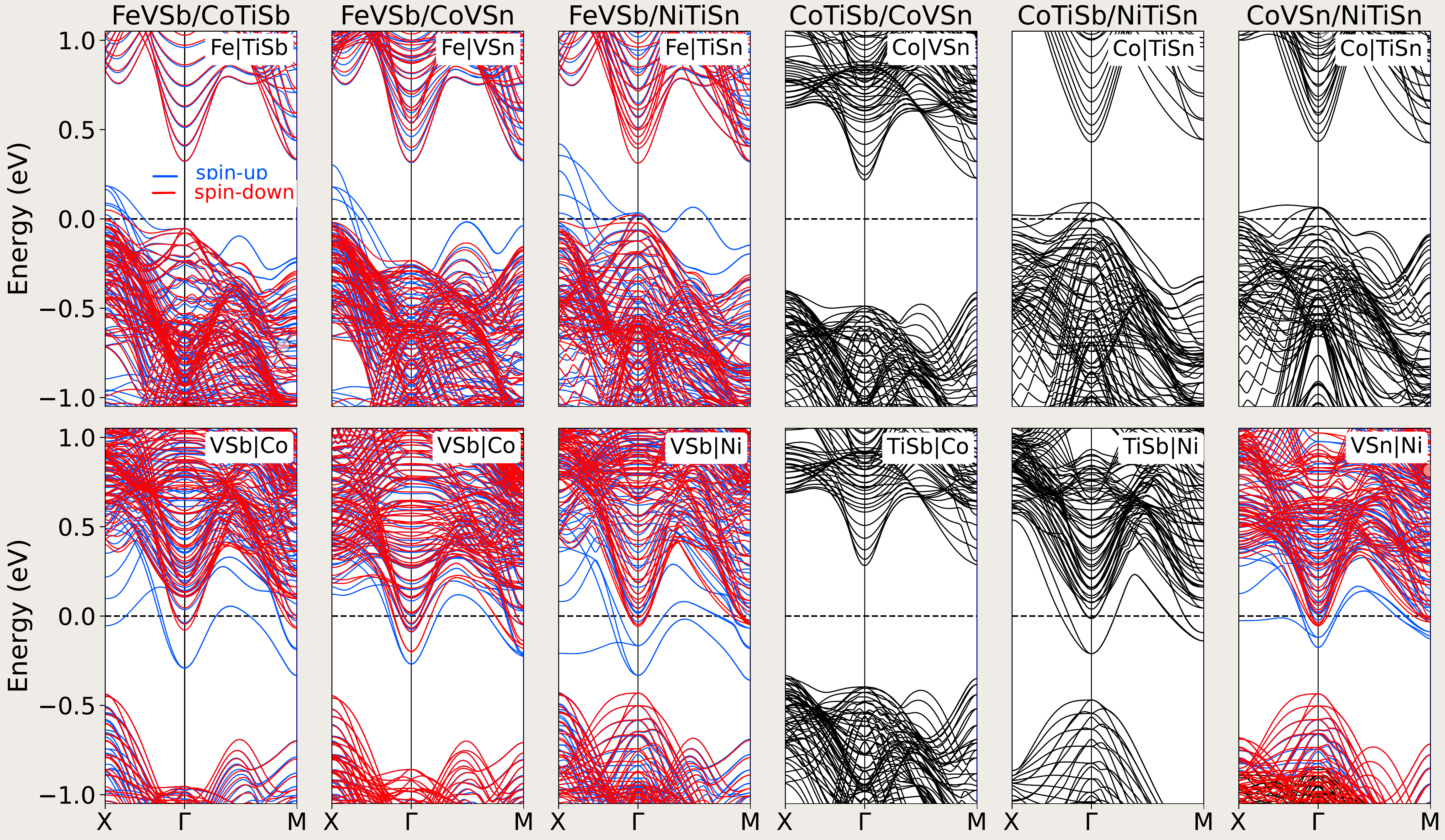}
\caption{Band structure for all considered heterojunctions along the X-$\Gamma$-M high symmetry 
directions of the 2D Brillouin zone.}
\label{fig4}
\end{figure*}

The study of the bulk systems is followed by the calculation of the interface properties of
heterojunctions. First, in Fig.\,\ref{fig3}, we present the band alignment of all considered 
heterojunctions. Our four compounds result in six possible heterojunctions and among them only 
FeVSb/CoTiSb and CoTiSb/NiTiSn possess type-I band alignment meaning that the CBM of FeVSb (NiTiSn) 
is higher than the CBM of CoTiSb, and the VBM of both FeVSb and NiTiSn is lower in energy than 
the VBM of CoTiSb. All other four heterojunctions are of Type-II character. Note that 
the band alignment of the heterojunctions is calculated according to the procedure presented 
in Ref.\,\cite{Sharan2019}. We have simulated the interfaces as discussed in detail in 
Section\,\ref{sec2} and in Table\,\ref{table2} we summarize all our results. First, we should 
note that for each heterostructure, there are two possible terminations at the interface denoted 
as Termination-1 when the first compound ends at a pure X layer and Termination-2 when it ends 
at a mixed YZ layer. As we discussed above in Section\,\ref{sec2} the second compound adopts 
the in-plane lattice constant of the first compound, and its out-of-plane lattice constant 
changes accordingly in order to preserve the unit cell volume as shown by the $c/a$ ratios 
presented in Table\,\ref{table2}.

With the exception of the CoTiSb/CoVSn case, in all other studied heterojunctions, the number of 
valence electrons at the interface is no more 18 but there is either an excess of electrons 
($Z_\mathrm{t}$ taking into account the two layers which form the interface is larger than 
18) or a deficit of electrons ($Z_\mathrm{t}$ is smaller than 18) and the interface is metallic. 
In the first case of electron excess, we should have the creation of a 2DEG, while electron 
deficit can be translated to an excess of holes leading to 2DHG. But as the band structures 
projected on the (001) plane and presented in Fig.\,\ref{fig4} reveal the situation is more 
complex. In most cases, the metallic interface is also magnetic and the 2DEG (2DHG) is
spin-polarized. In Table\,\ref{table2} we present also the spin magnetic moments of the atoms 
at the two interface layers as well as the total spin magnetic moment taking the sum  of 
the spin moments of all interface atoms.

\begin{figure}[!b]
\centering
\includegraphics[width=\columnwidth]{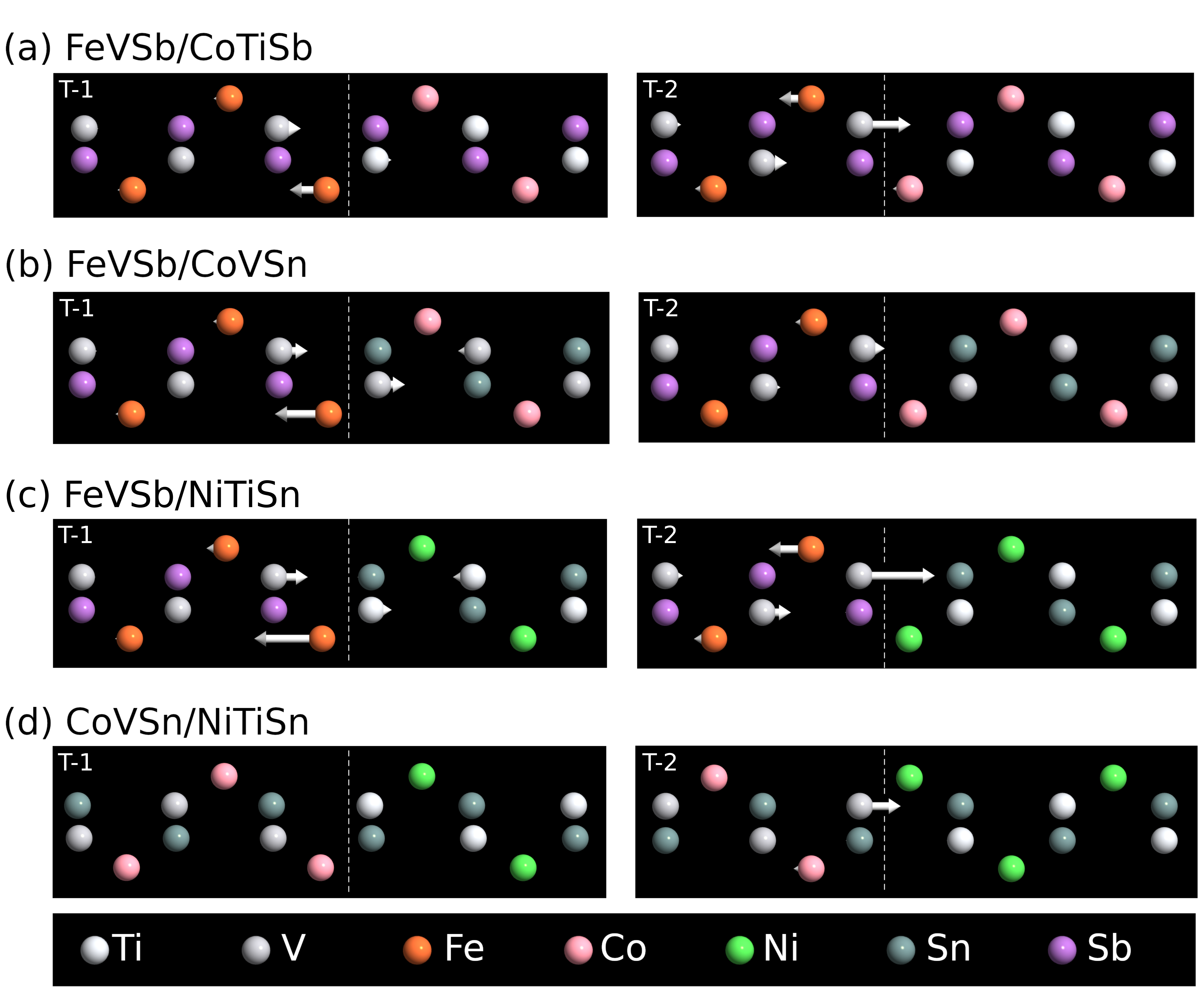}
\caption{(Color online) For both possible interface structures (denoted as T1 and T2) 
for each system, we show the structure of the interfaces. Arrows denote the direction of the 
spin magnetic moments of the atoms at the interface and their magnitude is proportional to 
the absolute value of the atomic spin magnetic moments presented in Table\,\ref{table2}.}
\label{fig5}
\end{figure}

The heterojunctions where FeVSb is one of the two junction materials are the most interesting 
cases. When Fe is at the interface, there is a deficit of electrons and we have a 2D spin-polarized 
hole gas (SPHG). In the band structures shown in Fig.\,\ref{fig4} this is reflected in the 
conduction bands, which are crossed by the Fermi level. When the interface layer is VSb and 
not Fe, there is an excess of electrons leading to 2D spin-polarized electrons gas (SPEG) and 
now the Fermi level crosses the valence bands. Moreover, the Fe and V atoms of FeVSb at the 
interface and subinterface layers are the ones responsible for the spin-polarized character of the
electron (hole) gas. This is depicted clearly in Fig.\,\ref{fig5}, where we present a 
schematic representation of the layers around the interface for all three heterojunctions
containing FeVSb and for both terminations. Arrows show the direction of the atomic spin 
magnetic moments and their magnitude is proportional to the values of the spin magnetic moments. 
In all three heterojunctions presented in the figure, it is easily observed that the magnetic moments 
reside primarily at the Fe and V atoms at the interface and subinterface layers irrespective of 
whether we have Fe (Termination-1) or VSb (Termination-2) interface layers.

\begin{figure*}[t]
\centering
\includegraphics[width=\textwidth]{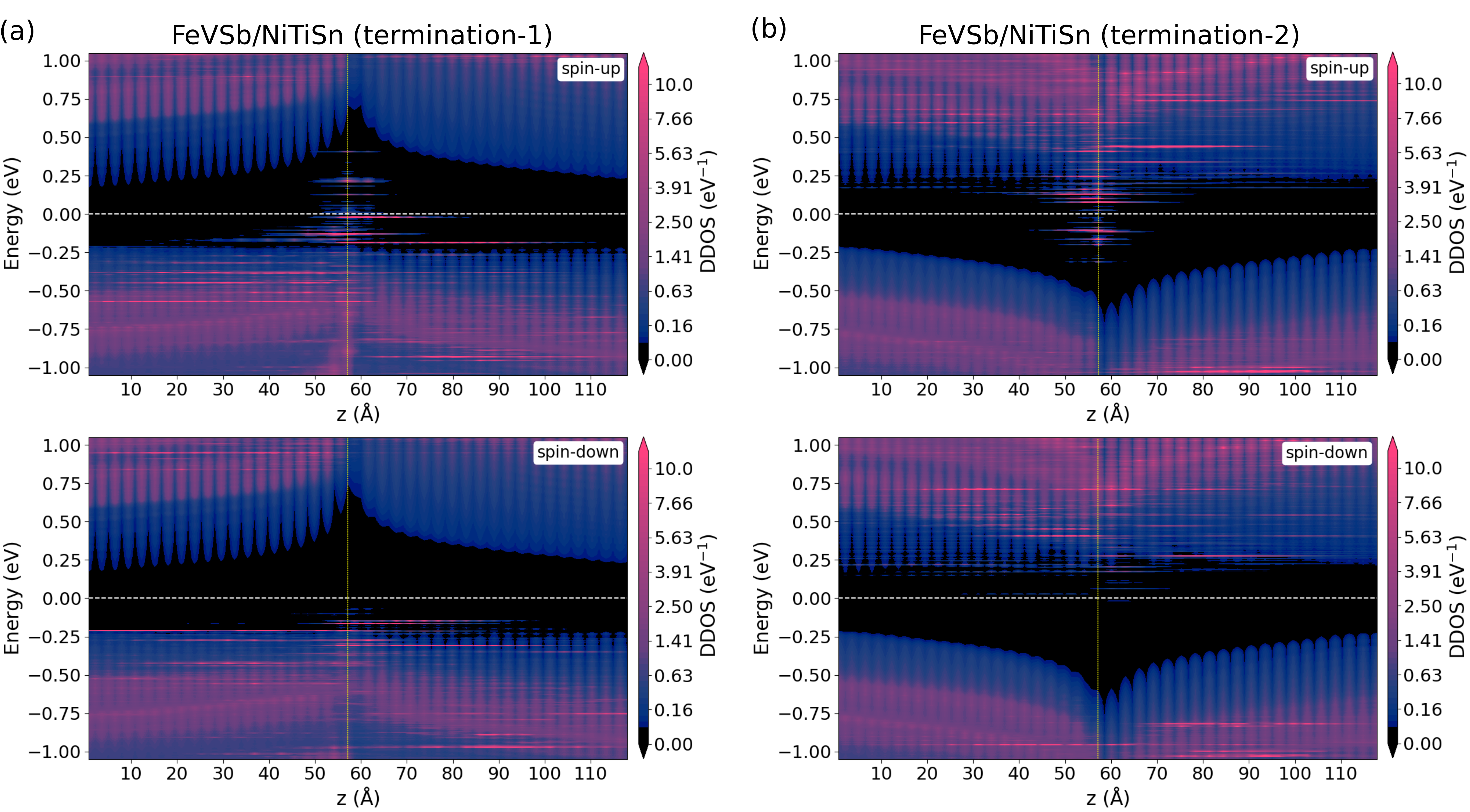}
\vspace{-0.5 cm}
\caption{(a) (Color online)  Projected device density of states (DDOS)  for the 
spin-up (upper panel) and spin-down electrons (lower panel) for the first termination 
of the FeVSb/NiTiSb junction (the atomic structure for the interface region  is given 
in \ref{fig5}). The white dashed lines display the Fermi level while the vertical 
yellow dashed lines denote the interface. (b) The same as (a) for the second termination.}
\label{fig6}
\end{figure*}

In the case of Termination-I interfaces, the Fe atoms at the interface layer carry sizeable 
magnetic moments which are antiparallel to the spin magnetic moments of the Ti (V) interface atoms as 
shown in Table\,\ref{table2} and schematically in Fig.\,\ref{fig5}. In the case of Termination-2 
interfaces the Co (Ni) atoms carry very small spin magnetic moments and it is the V atom at 
the interface layer of FeVSb, which carries again the main portion of the spin magnetic moment.
CoVSn/CoTiSb is a particular case due to the presence of Co in both materials of the heterojunction, 
and the resulting interface remains semiconducting for both terminations (see Fig.\ref{fig4}). 
In the case of CoTiSb/NiTiSn interfaces the interface is metallic but non-spin-polarized and thus 
we have either a usual 2DEG or 2DHG behavior at the interface as shown in Table\,\ref{table2} and as 
can be deduced from Fig.\,\ref{fig4}. Finally, in the case of the CoVSn/NiTiSn, termination-II 
is spin-polarized and the 2DEG is also spin-polarized, while termination-I is simply metallic, 
as deduced also from the band structure in Fig.\,\ref{fig4}, and there is a usual 2DHG at the 
interface.

To confirm our conclusions we have also performed device calculations for both possible 
terminations in the case of the FeVSb/NiTiSn heterostructure. As discussed in Section\,\ref{sec2} 
device calculations are much more demanding in computer resources than the supercell 
calculations presented up to now. In this case, we have a 120\,\AA\ heterojunction 
(i.e., the scattering region) with two semi-infinite leads. The obtained spin magnetic 
moments are similar to the ones obtained using the supercell approach shown in Table\,\ref{table2}. 
In Fig.\,\ref{fig6} we present for both terminations the spin-up and the spin-down 
device density of states (DDOS) as a function of the distance. For the spin-up DOS 
(upper panels) there is a finite DOS around the interface layers which quickly vanishes 
as we move away from the interface. In the case of the spin-down DOS for both 
terminations (lower panels) there is a negligible DOS around the Fermi level around the
interface region of the device. Thus at the interface, we have a nearly half-metallic 
magnetic behavior and as we move away from the interface we get non-magnetic semiconducting 
behavior.

The results discussed in the previous paragraph agree well with the conclusions
in Ref.\,\cite{Sharan2019} regarding the confinement of the electrons and holes at 
the interface. In Ref.\,\cite{Sharan2019} the authors have studied the case of CoTiSb/NiTiSn 
heterojunction. At the TiSb-Ni interfaces, which accommodates a 2DEG, electrons are 
confined in a region of about 1.5\,nm (15\,\AA) around the interface layer. At the Co-TiSn 
interface, there is a 2DHG and the holes are confined in a slightly larger region of 
about 20\,\AA\ around the interface. In our case of FeVSb/NiTiSn heterojunction, 
as shown in Fig.\,\ref{fig6}, the change in the DDOS around the  interface layer occurs 
in a region of about 20\,\AA\ around the Fermi level in the case of the Fe-TiSn interface 
and a slightly thicker region in the case of the VSb-Ni interface. Thus the electron 
and  hole gas are confined in a similar region as in Ref.\,\cite{Sharan2019}.

\begin{figure*}[!ht]
\centering
\includegraphics[width=\textwidth]{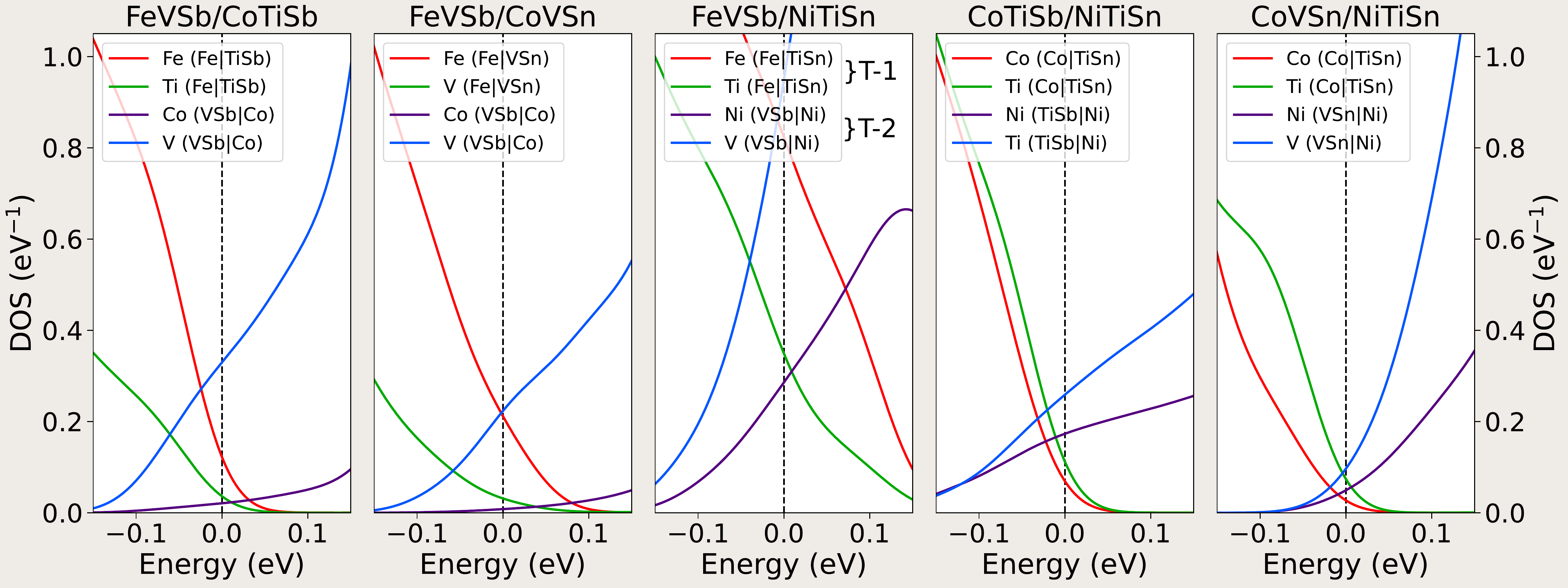}
\vspace{-0.4 cm}
\caption{Atom-resolved non-magnetic (paramagnetic) total density of states (DOS) of the transition 
metal atoms at the two interface layers around the Fermi level of the heterojunctions.}
\label{fig7}
\end{figure*}

\subsection{Origin of the spin-polarization at the interface}

The emergence of interface spin polarization can be  qualitatively explained on the basis 
of the Stoner model for itinerant ferromagnetism. In materials where the Stoner criterion is fulfilled 
($I\cdot N(E_F) > 1$, where $I$ is the Stoner parameter and $N(E_F)$ is the non-spin-polarized 
DOS at the Fermi level), the occurrence of magnetism is favored. To examine whether the Stoner 
model can be applied to the interfaces one should first estimate the value of $I$ and $N(E_F)$. 
For the latter in Fig.\,\ref{fig7} we have plotted the non-spin-polarized DOS around the Fermi 
level for the atoms at the interface of heterojunctions except the CoTiSb/CoVSn one which is 
a non-magnetic semiconductor. One can easily  see that the FeVSb/NiTiSn heterojunction for 
both terminations presents a significantly larger DOS at the Fermi level than all other cases 
which approach the value of 1 eV$^{-1}$.

To estimate the value of the Stoner parameter $I$, we can use the relationship $I=\frac{U + 6J}{5}$
proposed by Stollhoff \textit{et al.}, where $U$ is and $J$ are the Hubbard on-site Coulomb repulsion 
and the exchange parameters, respectively. These parameters are hard to extract experimentally and 
their \textit{ab-initio} calculation is very demanding. In Ref.\,\cite{Sasioglu2013}, their values 
have been calculated for several Heusler materials including also some half-Heusler compounds 
using the constrained random-phase-approximation. For the transition metal atoms of half-Heusler 
compounds the $U$ parameters calculated in Ref.\,\cite{Sasioglu2013} vary between 3 eV and 4 eV and 
the $J$ parameter is around 0.7 eV. We expect that since these values are for bulk systems, in the 
case of 2D systems like the ones studied here, the $U$ values will be slightly larger due to reduced 
screening stemming from  the out-of-plane atoms in the heterojunctions. Thus, using the $U$ and 
$J$ values from the Ref.\,\cite{Sasioglu2013} the relation mentioned above gives for the $I$ parameter  
a value of 1 eV to 1.4 eV. According to this, we should expect that only the two terminations of 
the FeVSb/NiTiSn heterojunction should present a spin-polarized electronic band structure at the 
interface. But the Stoner model for itinerant magnetism is a mean-field treatment missing the 
effect of strong electronic correlations. The latter can induce also magnetic order \cite{Durr1997} as 
it seems to be the case for all the other interfaces which present a low DOS at the Fermi level as 
shown in Fig.\,\ref{fig6} and do not fulfill the Stoner criterion.

\subsection{Modified Slater-Pauling Rules for Interfaces}

Slater-Pauling rules in the case of Heusler compounds were initially formulated 
in the case of half-metallic Heuslers crystallizing in the C1$_\mathrm{b}$ lattice 
as the one adopted by the present compounds \cite{Galanakis2002a}. These rules 
connect the half-metallicity and the magnetic properties in the case of Heusler 
compounds. It was shown in Ref.\,\cite{Galanakis2002a} that the total spin magnetic 
moment in the unit cell $M_\mathrm{t}$ in $\mu_\mathrm{B}$ units is just the total 
number of valence electrons in the unit cell, $Z_\mathrm{t}$ minus 18 for half-metals 
($M_t=Z_t-18$). This rule expresses the fact that in the spin-down band structure, where 
the energy gap exists, there are exactly 9 completely occupied bands. When $Z\mathrm{t}$ 
equals 18 the total spin magnetic moment is zero. Since in the Heusler compounds crystallizing 
in the C1$_\mathrm{b}$ lattice, conventional antiferromagnetism cannot occur due to 
symmetry reasons, the 18 valence electron Heusler compounds have to be semiconductors
or completely compensated ferrimagnetic half metals. This prediction is in agreement 
with the behavior of FeVSb, CoVSn, CoTiSb, and NiTiSn compounds.

\begin{figure}[t]
\centering
\includegraphics[width=\columnwidth]{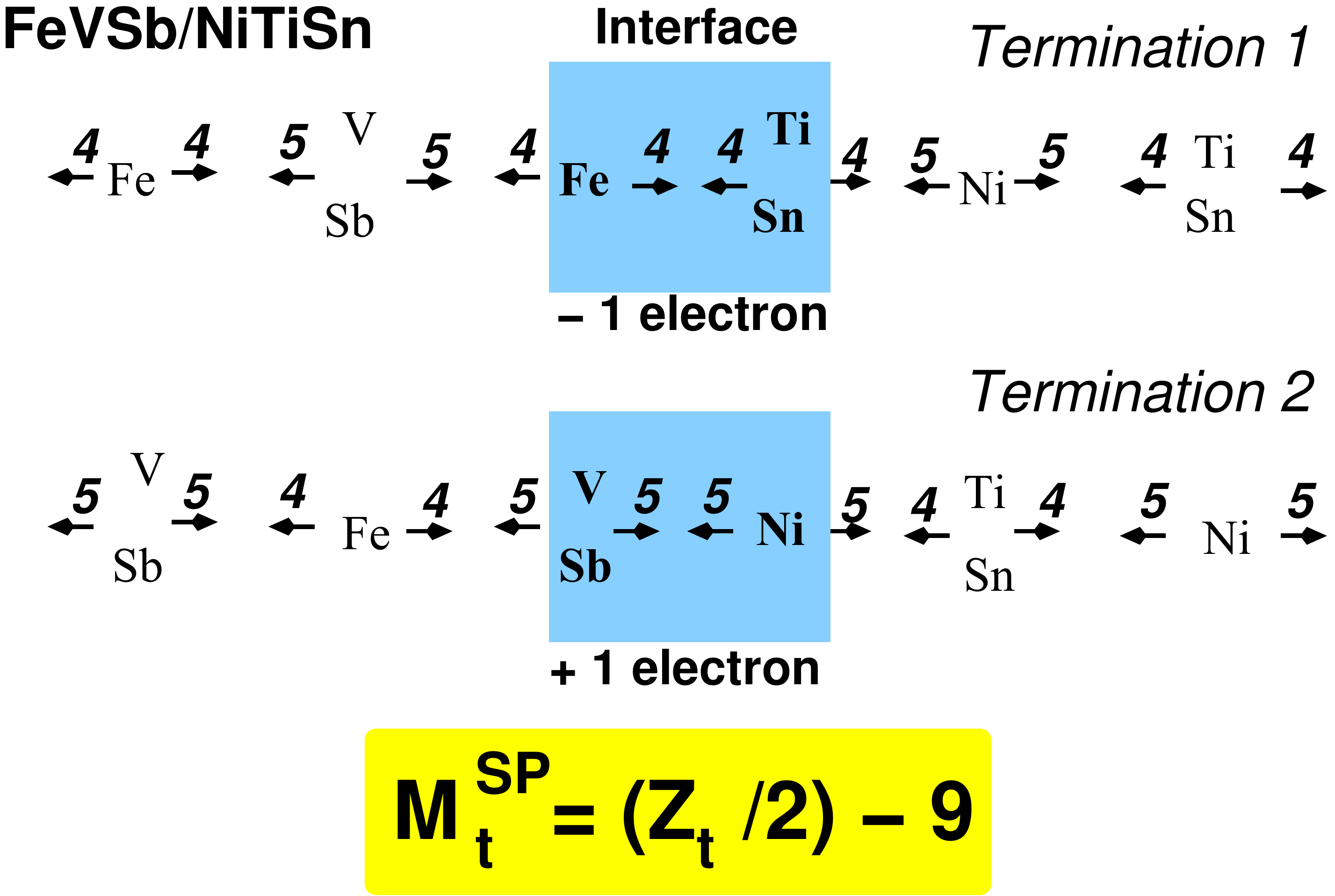}
\caption{Schematic representation of the bonding in the two possible FeVSb-NiTiSn 
interfaces. For the half-metallic materials  from
both sites of the interface the bonds between the atoms at two consecutive 
layers are built up from 9 valence electrons. The numbers on top of the arrows show the 
valence electrons contributed by each layer (e.g. VSb has ten valence electrons which are 
splitted for the bonds at the two sides of the VSb layer). At the interface, the electrons 
contributing at the bonding are $\frac{Z_{\mathrm{t}}}{2}$ (see Table \ref{table2} for the 
definition). Thus the excess or shortage of valence electrons at the interface with respect 
to the nine within the rest of the material is $\frac{Z_{\mathrm{t}}}{2}-9$. To achieve half 
metallicity these extra(less) valence electrons should occupy(vacate) spin-up states leading 
to a modified Slater-Pauling rule for the total spin magnetic moment at the interface: 
$M_{\mathrm{t}}^{\mathrm{SP}}= \frac{Z_{\mathrm{t}}}{2}-9$ where $M_{\mathrm{t}}^{\mathrm{SP}}$ 
is in $\mu_\mathrm{B}$. }
\label{fig8}
\end{figure}

The question which rises is whether it is possible to formulate the Slater-Pauling rule 
in a way to connect the magnetic properties of the studied heterojunctions  
with the total number of valence electrons at the interface. To answer this question, 
one should consider the (001) planes of the lattice shown in Fig.\,\ref{fig1} and we will 
use FeVSb as the example.  The (001) planes are made up either of pure Fe or mixed V-Sb 
atoms. The FeVSb unit cell has in total 18 valence electrons; Fe contributes 8 valence 
electrons and the V-Sb atoms 10. In a simplified representation like the one shown in 
Fig.\,\ref{fig8}, each Fe plane contributes 4 electrons to the bonds with the VSb plane 
on its right and another 4 electrons to build up bonds with the VSb plane on its left. 
Following the same reasoning, each VSb plane contributes 5 electrons to each one of its 
two sides. Thus in total 9 electrons contribute to the bonding between the atoms in two 
consecutive layers. The same reasoning stands also for NiTiSn but now Ni planes contribute 
5 electrons to the bonds with each one of the neighboring TiSn layers and each TiSn layer 
contributes 4 electrons to the corresponding bonds. At the VSb-Ni or TiSn-Fe interfaces 
shown in Fig.\,\ref{fig7} we have now  10 or 8 electrons, respectively, contributing to 
the bonding between the atoms at the interface layers. The total number of these 
electrons is $Z_\mathrm{t}/2$, where $Z_\mathrm{t}$ refers to the compound built up from 
the atoms at the interface (\textit{e.g.} NiVSb or FeTiSn in our case) and thus the 
difference with the nine electrons of the perfect semiconducting interface is $Z_\mathrm{t}/2-9$. 
This means that at the VSb-Ni interface, there is a surplus of one electron, while at the 
TiSn-Fe interface, there is a shortage of one electron. Thus the prerequisite for 
half-metallicity at the interfaces is that the total spin magnetic moment at the interface 
follows a modified Slater-Pauling rule of the $M_\mathrm{t}=Z_\mathrm{t}/2-9$ form.

To verify the validity of the proposed modified Slater-Pauling rule in Table\,\ref{table2} 
we present the calculated total spin magnetic moment for all interfaces with respect 
to the ones predicted by the Slater-Pauling rule. For the discussion, we take into account 
the electronic character of the interface as presented in Fig.\,\ref{fig4} where the (001) 
projected band structures for all interfaces are presented. The calculated total spin 
magnetic moments are very close to the ones predicted by the Slater-Pauling rule for 
half-metallic behavior (deviation less than 0.05 $\mu_\mathrm{B}$) in the case where FeVSb 
is part of the heterojunction and Fe is at the interface as well as for the VSn-Ni interface. 
If we examine the band structures presented in Fig.\,\ref{fig4} the Fe-VSn interface is a 
perfect half-metal. In the case of Fe-TiSb interface, the Fermi level slightly crosses the 
spin-down conduction band, while in the case of VSb-Co and VSn-Ni interfaces, the Fermi level 
slightly crosses the spin-down conduction band. Both types of interfaces between the CoTiSb 
and CoVSn compounds are semiconducting in agreement with the Slater-Pauling rule. For the 
rest of the interfaces, the Slater-Pauling rule is not obeyed  and the electronic character 
of the interfaces is no more half-metallic.  Magnetism in materials appears when the gain 
in energy due to the lowering of the DOS at the Fermi level (fewer electrons with the 
maximum energy) overcomes the cost in band energy (energy to flip spins and create the
spin imbalance needed for magnetism). In half-metallic magnets, the magnitude 
of both competing energy contributions increases but the final gain in energy is larger than 
in the simple magnetic case. Otherwise, the material remains  magnetic without being half-metallic 
explaining the different behavior of the interfaces under study.

\begin{figure}[!b]
\centering
\includegraphics[width=\columnwidth]{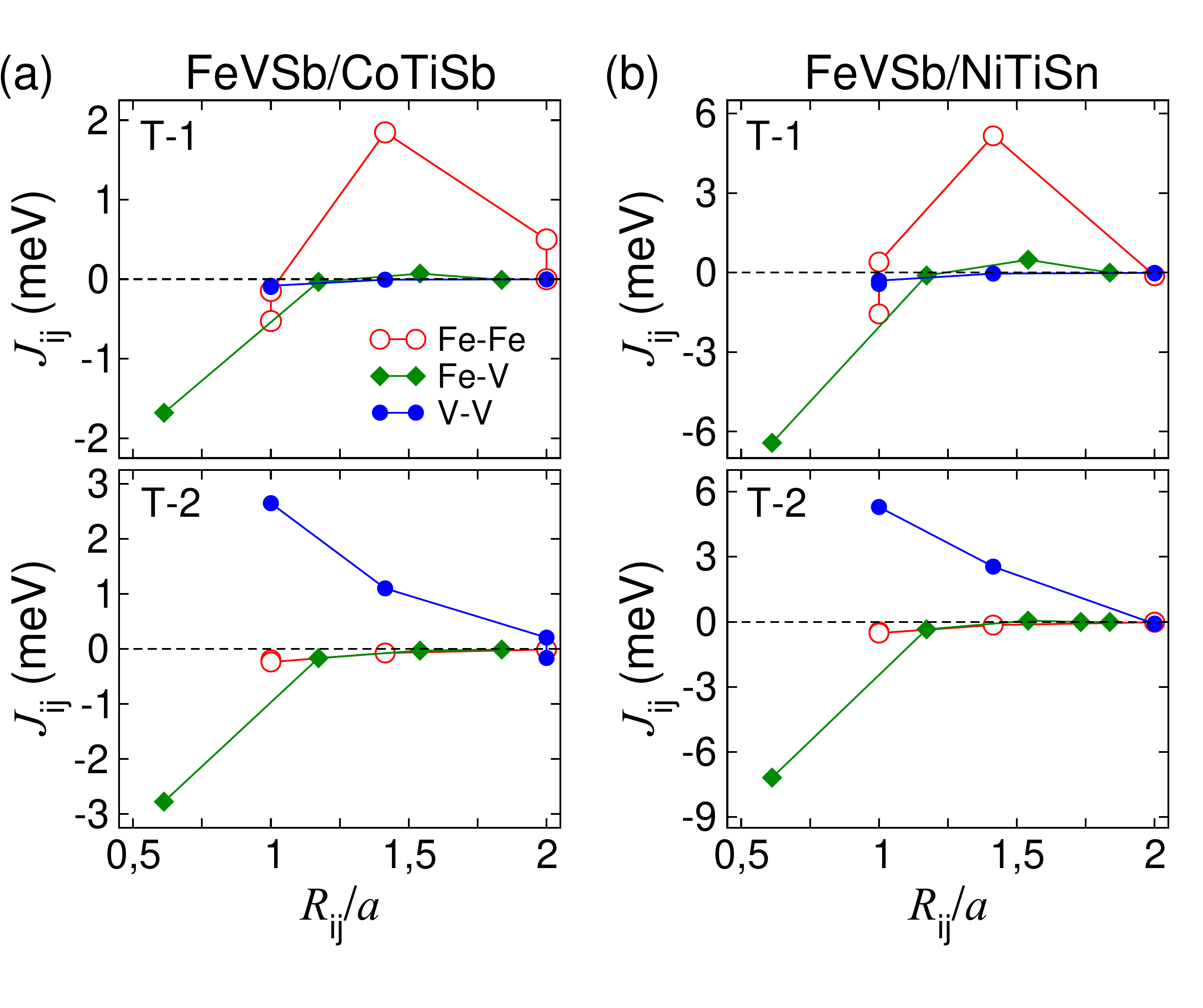}
\vspace{-0.4 cm}
\caption{(a) (Color online) Calculated intra-sublattice (Fe-Fe and V-V) and 
inter-sublattice (Fe-V) Heisenberg exchange parameters as a function of distance
for the interface and sub-interface layers of the FeVSb/CoTiSb heterojunction for
both possible T-1 and T-2 interface terminations. (b) The same as (a) for the 
FeVSb/NiTiSn heterojunction. For the definition of T-1 and T-2 see Figs.\,\ref{fig1} and 
\ref{fig5}.}
\label{fig9}
\end{figure}

\subsection{Exchange Interactions and Curie Temperature}

In the last part of our study, we compute the Heisenberg exchange parameters as discussed 
in detail in Section\,\ref{sec2}. We have chosen the FeVSb/CoTiSb and FeVSb/NiTiSn heterojunctions 
since both follow the modified Slater-Pauling rule described above. We present our results 
in Fig.\,\ref{fig9}. The upper panels refer to Termination-1, where the interface is Fe-TiSb 
(TiSn) and the lower panels to Termination-2, where the interface layers are VSb-Co(Ni). We 
restrict ourselves to FeVSb  only, since only the Fe and V atom  at the interface present significant 
spin magnetic moments, and present the exchange constants as a function of the distance for the 
Fe and V atoms at the interface and subinterface layers. The first striking characteristic feature in 
all cases is the strong negative exchange parameters between the nearest neighbor Fe-V atoms. 
This is reflected in the antiparallel Fe-V spin magnetic moments which stabilize the magnetic 
order at the interface. In the case of Termination-1, the V atoms are at the subinterface layer, 
the V has negligible spin magnetic moments and the V-V exchange constants are vanishing. In the 
case of Termination-2, the V atoms are at the interface layer and they carry a sizeable spin 
magnetic moment as shown in Table\,\ref{table2} (see also Fig.\,\ref{fig5}) and the V-V exchange
interactions are all positive, favoring the ferromagnetic alignment of the V spin magnetic moments. 
For Termination-2, the interactions between the Fe atoms located at the subinterface layer are negligible as expected. Interestingly in the Termination-1 cases, the Fe-Fe exchange parameters 
are sizeable for both  FeVSb/CoTiSb  and FeVSb/NiTiSn heterojunctions favoring the ferromagnetic 
alignment of the Fe spin magnetic moments at the interface layers.

\begingroup
\begin{table}[t]
\caption{Calculated magnetic anisotropy energy (MAE) and the interface 
Curie temperature $T_C$ for the FeVSb/CoTiSb (NiTiSn) heterojunctions for both
possible interface terminations. }
\begin{ruledtabular}
\begin{tabular}{c|c|c|c}
Half-Heusler &  Interface  & MAE=(E$_{||}$-E$_{\perp}$)  & $T_\mathrm{C}$ (K) \\ 
heterojunction     & termination             &  ($\mu$eV/cell)    &    \\
\hline
 FeVSb/CoTiSb & Fe|TiSb & -121.3   &  78  \\
              & VSb|Co &  -76.6   & 101  \\ 
       
FeVSb/NiTiSn & Fe|TiSn  & 8.5   &  187    \\
             & VSb|Ni   & -24.9  &  248 \\

\end{tabular}
\end{ruledtabular}
\label{table3}
\end{table}

We used the calculated exchange parameters to estimate the Curie temperature $T_\mathrm{C}$ 
within the mean-field approximation (MFA) and present our results in Table\,\ref{table3}. 
The obtained values range from about 78 K to about 248 K being larger for the FeVSb/NiTiSn 
heterostructure and for Termination-2 with respect to Termination-1. These values are 
relatively small when compared to room temperature. Unfortunately, no experimental data 
on the Curie temperatures exist in the literature for comparison to establish the accuracy of 
our calculations. Such a comparison would be necessary since (i) MFA ignores spin-fluctuations 
which are important especially for low-dimensional magnets tending to overestimate $T_\mathrm{C}$, 
and (ii) for systems with small spin magnetic moments (less than 1 $\mu_\mathrm{B}$),
the exchange parameters $J_{ij}$ are underestimated when using the linear response theory 
(as is the case here) and thus the $T_\mathrm{C}$ is also underestimated \cite{Pajda2001}.
These two phenomena induce competing errors and this explains the behavior of calculated 
$T_\mathrm{C}$ in bulk half-ferromagnetic Heusler compounds; in the case of NiMnSn  where 
the Mn spin magnetic moment is very large the MFA  overestimates the $T_\mathrm{C}$ by 
about 400 K, while in Co$_2$CrAl and  Co$_2$MnSi full-Heusler compounds due to the much 
smaller spin magnetic moments,  MFA is much more accurate and underestimates $T_\mathrm{C}$ 
by about  50-100 K only \cite{Sasioglu2005}. In the case of the studied interfaces due 
to the  small spin magnetic moments at the interface, we expect that the errors induced 
by the two competing phenomena will almost cancel out each other and the agreement with 
the experiment will be even better than in full-Heusler compounds.

Finally, in Table.\,\ref{table3} we also present the calculated values of the magnetic 
anisotropy energy (MAE) for FeVSb in $\mu$eV per cell. All systems possess small values 
of MAE of the order of 0.1 $m$ eV.  In all cases, the easy magnetization axis is perpendicular 
to the interface with the exception of Termination-1 in the FeVSn/NiTiSn system where the 
easy magnetization axis is parallel to the interface but the MAE is almost vanishing. As 
discussed above when we considered our supercells, FeVSb adopted in all studied cases a 
cubic lattice and we relaxed the cell only of the other material in the heterojunction. 
This explains the very  small calculated MAE values. If the vice-versa procedure takes 
place and  now FeVSb is grown on top of the other material (\textit{e.g} CoTiSb), the  
latter will adopt a cubic lattice and FeVSb will adopt the same in-plane lattice constant. 
This will lead to a tetragonal structure for the FeVSb material and much larger MAE 
values \cite{Aull2021}, since materials tend to keep their unit cell volume almost constant 
and change accordingly their lattice parameters.

\section{Summary and Conclusions}\label{sec4}

Half-Heusler compounds, which have 18 valence electrons per unit cell, like FeVSb, CoTiSb, 
CoVSn and NiTiSn are well-known for their non-magnetic semiconducting properties. 
In the present study, we employed first-principles electronic band structure calculations 
and examined the properties of the interfaces of the heterojunctions based on  these four 
half-Heusler compounds. First, we confirmed the non-magnetic semiconducting character of these 
compounds in their bulk form. Then we used a supercell approach to study the heterojunction
interfaces considering the [001] axis as  the growth direction. Our results showed that 
several of these interfaces become metallic and in several cases also magnetic. The emergence 
of spin polarization is accompanied by the formation of a two-dimensional electron (2DEG) 
or hole (2DHG) gas at  the interface making such structures promising for spintronic applications. 

To further understand the magnetic properties of the interfaces of heterojunctions, we developed 
a modified Slater-Pauling rule that is similar to the corresponding rule for the bulk half-metallic 
half-Heusler compounds. This rule connects the altered number of valence electrons at the interface 
to the total spin magnetic moment at the interface and is a prerequisite for half-metallicity to 
occur. We also calculated other properties of interest, such as exchange parameters, Curie 
temperature, and magnetic anisotropy energies.

Overall, our study adds to the growing body of knowledge on the properties of half-Heusler
heterojunction interfaces and their potential for use in spintronic and magnetoelectronic 
devices. We hope that our findings, combined with recent experimental evidence for the 
presence of 2DEG at  CoTiSb/NiTiSn heterojunctions \cite{Harrington2018}, will increase 
interest in these  materials and their potential applications. We expect our study to motivate 
future efforts and studies  toward the experimental realization of devices using the proposed 
heterojunctions.

\acknowledgements

This work was supported by SFB CRC/TRR 227 of Deutsche Forschungsgemeinschaft (DFG) 
and by the European Union (EFRE) via Grant No: ZS/2016/06/79307. B. S. acknowledges 
financial support from Swedish Research Council (grant no. 2022-04309). The computations 
were enabled in project SNIC 2021/3-38 by resources provided by the Swedish National 
Infrastructure for Computing (SNIC) at NSC, PDC, and HPC2N partially funded by the 
Swedish Research Council (Grant No. 2018-05973). B.S. acknowledges allocation of 
supercomputing hours by PRACE DECI-17 project `Q2Dtopomat' in Eagle supercomputer 
in Poland and EuroHPC resources in Karolina supercomputer in Czech Republic.

\providecommand{\noopsort}[1]{}\providecommand{\singleletter}[1]{#1}%

\end{document}